\documentclass[pre,floatfix,twocolumn,superscriptaddress,nofootinbib]{revtex4-2}
\usepackage{graphicx}
\usepackage[USenglish]{babel}
\usepackage{float,color}
\usepackage{lipsum}
\usepackage{epigraph}

\begin{document}

\title{Broken Symmetries, Information and Emergence: What is theory, 
that biology should be mindful of it?}

\author{Shakti N. Menon}
\affiliation{The Institute of Mathematical Sciences, CIT Campus, Taramani, Chennai 600113, India}
\author{Sitabhra Sinha}
\affiliation{The Institute of Mathematical Sciences, CIT Campus, Taramani, Chennai 600113, India}
\affiliation{Homi Bhabha National Institute, Anushaktinagar, Mumbai 400094, India}

%
%

\begin{abstract}
The discipline of `theoretical biology’ has been developing from its inception several decades ago almost in parallel with the advances in biology, so much so that the latter is often considered to be almost exclusively an empirical science. However, the scenario has been changing in recent years with statistical mechanics, nonlinear dynamics and soft-matter physics being more and more frequently invoked to explain various biological observations. As distinct from computational biology, theoretical biology is not just an attempt to reproduce \textit{in-silico} experimental phenomena, but asks more general and abstract questions. It strives to attain a more fundamental understanding of the mechanisms underlying biological phenomena, ranging from oscillations to strategic actions, that can be unified through the perspective that views organisms as processing information to respond appropriately despite the noise in their environment. We show through a number of investigations carried out by our group, on the emergence of systems-level phenomena through interactions between components, how an approach melding physics, and the theory of information \& computation can act as an unifying framework for biological processes across a wide range of temporal and spatial scales. 
\end{abstract}
\maketitle

\epigraph{It is an acknowledged truth in philosophy that a just theory will
always be confirmed by experiment. Yet so much friction, and so many
minute circumstances occur in practice, which it is next to impossible
for the most enlarged and penetrating mind to foresee, that on few
subjects can any theory be pronounced just, till all the arguments
against it have been maturely weighed and clearly and consistently
refuted.}{Thomas Malthus, ``An Essay on the Principle of
Population'' (1798)}

Can life -- the collection of processes associated with living organisms -- be understood using the explanatory framework often referred to as ``theory''? Our use of the term theory refers to a logically consistent set
of propositions that provide a general explanation for empirical observations and also implies certain consequences that can be tested experimentally. Although this approach, which involves an intricate interplay between observations/experiments and mathematical modeling, has proved so successful for physics over the preceding two centuries, even as recently as the 1980s the answer from persons of eminence in the life sciences, such as the evolutionary biologist Ernst Mayr~\cite{Mayr1982}, would have been an unequivocal ``No''~\footnote{Indeed, Mayr's last book, \textit{What Makes Biology Unique? Considerations on the Autonomy of a Scientific Discipline}, published in 2004, insists that ``biology, even though it is a genuine science, has certain characteristics not found in other sciences'' thus suggesting that physics (and chemistry) would be of little help in understanding living systems.}. One wonders how they would react to the vastly altered situation today, when publications in high-profile journals investigating biological processes not only report experimental observations, but are very often accompanied by results from mathematical modeling and computer simulations of the phenomena observed, with collaborations between biologists and researchers trained in physics more or less being the norm.

Although the trend of physicists gravitating toward questions related to the life sciences can be traced back to the beginning of molecular biology in the mid-twentieth century (with several of the pioneers of the field, most notably Crick and Delbr\"{u}ck being physicists by training), the formal division between the physical and life sciences is actually of rather recent origin. Aristotle, arguably the founding father of biology (at least in the European tradition) also wrote extensively on physics~\cite{Leroi2014}. The distinction he made between inanimate and the animate had to do with the latter having additional properties such as metabolism, ability to reproduce, sensory perception, motility and intellect -- properties that today we would call `emergent'', i.e., arising from the physico-chemical interactions between (inanimate) constituents. In fact, in his book on \textit{Physics}, Aristotle compares some of the characteristic animal functions to the workings of \textit{ta automata t\^{o}n thaumat\^{o}n} (loosely translated as automatic puppets). In particular, he refers to such automata in his discussions of two of the defining features of living organisms, viz., mobility and capacity to reproduce~\cite{Berryman2003}.

At the onset of the Age of Enlightenment in Europe, the mathematician and philosopher Ren\'{e} Descartes proposed that the body of a living organism is essentially a machine~\cite{Descartes}. In making such claims, he was likely influenced by his familiarity with mechanical devices, often animated by intricate clockwork mechanisms, that imitated the bodily movement of animals or men, and thus created an illusion of self-propelled actions -- one of the hallmarks of life~\cite{Chapuis1958}. The clockwork analogy reached its apogee with the dissemination of Newton's work on the laws of motion and gravitation, when classical mechanics came to be regarded as the ultimate explanatory framework and it was expected that the universe -- including life in it -- evolved according to deterministic dynamics given a specific set of initial conditions. From then on, the dominant approach used for explaining biological phenomenon tended to be provided by the most active branch of physics at the time. 

Thus, at the close of the $18^{\rm th}$ century, the initial growth of forays into understanding electricity naturally led to identifying electrical phenomena as the motivating force for life, following Luigi Galvani's demonstrations of how muscle movement in dead organisms can be elicited by electrical stimulation. Giovanni Aldini, Galvani's nephew, used this to dramatic effect by staging public spectacles in which bodies of recently executed convicts would be made to move (at times seeming to audiences that the corpses were trying to get up).\footnote{It is no coincidence that Mary Shelley's \textit{Frankenstein} (1818), which was written during this period, centers around the notion of electricity animating a corpse!} The porosity between the study of physical (i.e., non-living) and living processes continued as late as the first half of the $19^{\rm th}$ century, when physicians such as Thomas Young who were studying eyesight and color vision, eventually went on to make fundamental contributions towards establishing the wave aspect of light by experimenting with optical interference. The well-known Young's modulus was the result of another foray of the physician into physics -- this time trying to measure the elasticity of substances. It is quite likely that this could have been inspired by his well-known studies of hemodynamics in the blood circulatory system, whose veins and arteries need to have elastic properties for the flow to be unobstructed~\cite{Robinson2006}. The identification of biological sciences as a separate discipline around the middle of the $19^{\rm th}$ century was possibly the catalyst that eventually led to the cessation of dialogue with the physical sciences. This divide led to widespread belief that there existed an irreconcilable divergence between the properties of the animate and the inanimate (as evident, for example, in Mayr’s firm stance against the utility of mathematics and physics in the life sciences, alluded to earlier).

However, this notion began to be challenged by the beginning of the $20^{\rm th}$ century, when the pioneering Indian biophysicist (and also one of the inventors of wireless communication), Jagadish Chandra Bose showed that the response of living tissue such as muscles and plants to electrical stimulation bears a striking resemblance with the way inorganic substances, such as the magnetic oxide of iron, react to electricity~\cite{Bose1900}. As noted in a recent review~\cite{Dasgupta1998}, a receiver made of iron oxide (Fe$_3$O$_4$) was subjected by Bose to a single stimulus of electric waves, and the response was recorded in terms of the receiver's conductivity as a function of time. This curve resembled the response curve of muscle. The receiver was then exposed, in several experiments, to a train of such electrical stimuli. When the frequency was low, the distinction between each successive shock could be discerned. However, when the stimuli followed in rapid succession, the receiver could not recover from each individual `shock'. As has been noted, the ``effect may be described as \textit{tetanic}'', which refers to prolonged contraction of muscle caused by rapidly successive stimuli. Unfortunately Bose went on to claim that this was evidence that cosmic life permeated everything in the universe -- including inanimate objects. Unsurprisingly, this narrative did not conform to the modes of scientific communication and Bose's work was thus all but ignored by the scientific establishment.

The reentry of physics -- or more specifically, theoretical physicists -- into biology can be said to have begun in earnest around the 1940s, with some of the key pioneers of the quantum revolution in physics, such as Niels Bohr and Erwin Schr\"{o}dinger, turning their attention to the study of life. In an opening address titled ``Light and Life'' in a conference organized in 1932 at Copenhagen, Bohr suggested that the fundamental change in the way we view the natural world that had been brought about by quantum theory had made it imperative to look at the problems of biology from a fresh perspective~\cite{Bohr1933,Stent1989}. He went on to argue that the discovery by quantum physics of inherent limitations of a completely mechanical description of natural phenomena meant that the use of reductive explanations in biology needs to be reconsidered. Bohr also pointed out that the probabilistic nature of the world at the quantum level seemed to be at odds with the highly organized structure and function of living systems, even though all the physico-chemical processes underlying life must be subject to quantum principles. To this end, he provided the example of how the absorption of photons (whose behavior necessarily requires quantum mechanics to be understood properly) by photoreceptor cells in the eye, nevertheless leads to a deterministic response at the macroscopic level, viz., visual perception of its surroundings by the observer. 

Even more importantly, Bohr's interest in the relation between life and physics inspired one of his proteges, the physicist Max Delbr\"{u}ck to begin working on the effect of radiation on organisms~\cite{Fischer2007}. Following up on the discovery that X-rays can induce mutations in flies by the American geneticist Hermann Joseph Muller, Delbr\"{u}ck in collaboration with the Soviet Russian biologist Nikolay Timofeev-Ressovsky suggested that the molecular structure of genes must be such that perturbations produced by high energy radiation causes one stable state to switch to another. Explaining this stability, which contrasts with the statistical nature of molecular movements, was seen as the key puzzle in the yet to emerge area of ``molecular biology'', a term that was coined by the American scientist Warren Weaver in 1938~\cite{Judson1979}. The resulting publication by Timofeeff-Ressovsky, Zimmer and Delbr\"{u}ck influenced Schr\"{o}dinger, who popularized their ideas in his 1943 lectures at Trinity College, Dublin. These lectures marked a turning point in the understanding of cells and genes, and formed the basis of his book \textit{What is Life} that appeared the next year~\cite{Phillips2021}.
Many of the scientists who went on to establish the field of molecular biology cited Schr\"{o}dinger's suggestion that life, at the molecular level, was encoded in terms of an ``aperiodic crystal" capable of self-reproduction, as the primary stimulus that initially motivated them to move into the field. These included James D. Watson and Francis Crick, who elucidated the structure of the genetic material.

In a curious coincidence, the very same year, i.e., 1943, of Schr\"{o}dinger's lectures, saw the appearance of a paper ``A logical calculus of the ideas immanent in nervous activity" written by the American neurophysiologist Warren S. McCulloch and a self-taught math prodigy Walter Pitts. The paper showed that by simplifying neurons to just a 2-state device, viz., a switch that is at any time either ``Off'' (quiescent) or ``On'' (firing action potential), and by varying the threshold of the input (weighted sum of signals coming from other cells) beyond which the device switches states, small networks of neurons could be made to perform logical operations such as AND, OR and NOT. Therefore, the network could be viewed as implementing computation by processing propositions (signals) through logic gates (neurons). In the words of the AI pioneer Seymour Papert, through this work McCulloch and Pitts demonstrated that ``the laws governing the embodiment of mind should be sought among the laws governing information rather than energy or matter''~\cite{Papert1965}. This work proved to be the opening shot of the neural network revolution that we are currently in the middle of.

\begin{figure}[tbp]
   \includegraphics*[width=\columnwidth]{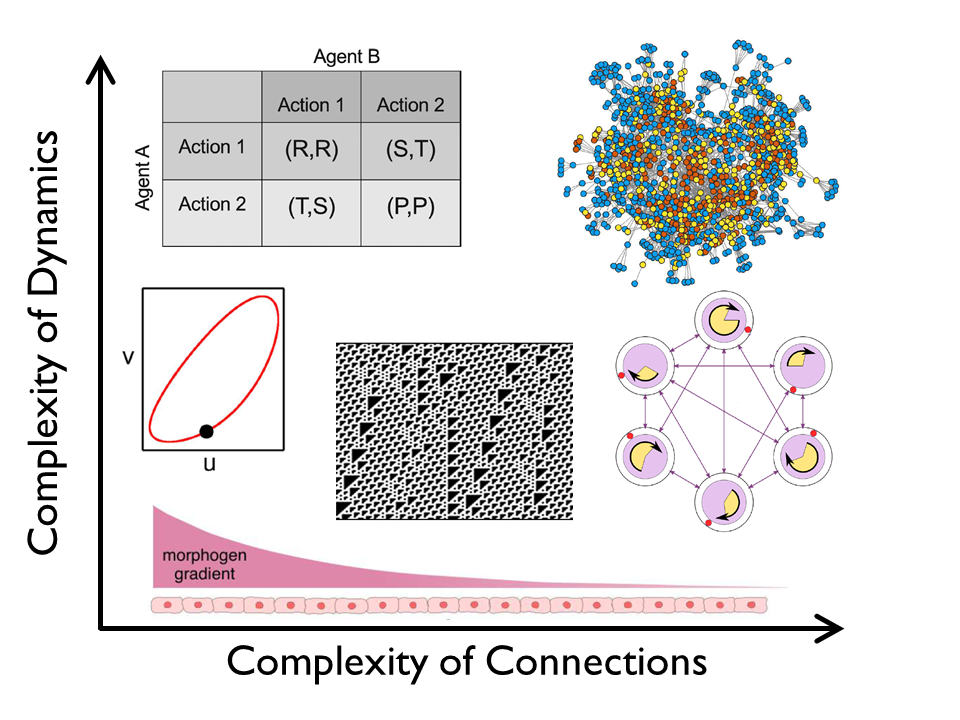}
   \caption{Schematic displaying a representative sample of systems of interacting units, classified according to the complexities of the underlying connection topologies, and that of the inherent dynamical processes associated with the individual units. These dynamics range from simple diffusion (bottom) and oscillations (middle left), to that of cellular
   automata (center), coupled phase oscillators (middle right) and 
   eventually, decision-making in games between rational agents (top left), who may be interacting with many others over a 
   complex social network (top right).}
      \vspace*{-2mm}
   \label{fig:fig1}
\end{figure}

These publications would go on to remold biology as a quantitative science whose central theme is \textit{information}. While in the case of molecular biology, the information is being transmitted from one generation to the next across time, for McCulloch and Pitts, it is being transmitted across space from one cell
to others that are connected to it in a complicated network.
In the context of development, cells require information about their spatial
position in the embryo so as to switch on appropriate genes that will allow 
appropriate cellular differentiation which is crucial for morphogenesis.
For instance, Wolpert invoked diffusing morphogens as the mechanism providing 
cells the cues to determine their locations relative to the boundary of the organism~\cite{Wolpert1969} (Fig.~\ref{fig:fig1}).
The information inherent in the resulting concentration gradient of morphogen molecules is used 
by the cells to follow spatially-dependent developmental trajectories~\cite{Kuyyamudi2022}.
Moving to a different scale, one can view contagia (either pathogens or
memes) as spreading across both space and time in a population of individuals,
connected to each other through kinship, professional, friendship or other
types of relations.
Thus, the key questions in the principal disciplines that study 
living processes, 
viz., evolutionary biology, ecology, developmental biology and neuroscience,
can be seen to be questions about how information is reliably received, stored, transmitted and interpreted/processed in order to enable an
organism to respond appropriately to its time-varying external and
internal environments.

%
That similar questions were being asked around the same time about processes involving very different spatial and temporal scales is probably not surprising if one considers the \textit{zeitgeist}. This era was permeated by ideas following from the pioneering work of Claude Shannon, as well as,
that of the group of scientists who were establishing the interdisciplinary field that came to be known as \textit{cybernetics}. Developed by McCulloch
and the American polymath Norbert Wiener (among others), 
it placed information as the key unifying theme across several sciences.
By also bringing out the importance of interactions between many components that result in qualitatively new behavior at the level of the 
entire system (Fig.~\ref{fig:fig1}), the field also prefigured one of the central concepts permeating contemporary physics-based approaches for investigating biological phenomena, viz., \textit{emergence}.
As brilliantly expounded by the condensed Philip W. Anderson in an article published more than a half century back~\cite{Anderson1972}, the phenomena observed at one scale cannot be simply described by 
invoking only the principles governing the actions of its components (observed at scales lower than that of the overall system), even though
the behavior of the latter is built upon the foundations of the former.
In Anderson's own words, ``At each stage entirely new laws, concepts, and
generalizations are necessary \ldots Psychology is not applied biology,
nor is biology applied chemistry''. The article also presciently indicated
that a general theory for understanding how bringing together many interacting units together (i.e., a quantitative change) leads to new
capabilities and behavior (i.e., a qualitative change)
would involve looking at `broken symmetries' as one moves from the level of the components to that of the entire system they comprise. 

The phenomenon of \textit{symmetry breaking}, i.e., the loss of invariance of an object or pattern under certain transformations -- such as translation,
rotation, reflection or scaling -- is of course ubiquitous in the biological world (see, e.g., Ref.~\cite{Marston2008}).
For instance, all organisms begin life as a single, round cell, the zygote, which is invariant under different transformations. However, over the
course of development, through various successive transitions, the 
embryo loses certain symmetries as it acquires the shape characteristic
of the organism. The loss of these symmetries, existing at the micro-level but disappearing at the macro-scale, are akin to the phase transitions in physical systems that distinguish one macroscopic state 
(characterized by ordering of a certain kind) from another (disordered) state. An obvious question that one can ask is where is the information
embedded that gives rise to the typical pattern that the system exhibits 
as a result of these transitions?

A facile answer would be to say ``It's in the genes''. However, genes
only encode proteins, and moreover every cell in a multi-cellular
organism has exactly the same set of genes (and thus, in effect, an identical
set of instructions). It is hence difficult to see how they help us understand
the \textit{bauplan}, the overall organizational structure of the organism,
resulting from phenotypically distinct, specialized cell types,
such as muscles or neurons, appearing in specific locations in the body
(e.g., in the limbs or in the head). Thus, the biochemical reactions
taking place in a living system must be somehow taking all the information
acquired from the DNA and from the environment (both that of
individual cells, as well as, that for the entire organism) and transforming
it into instructions that make morphogenesis possible. We can view
this information processing as \textit{computation}, where the transitions between
different states of the organism, each of whose information content can
be quantified (at least in principle), can be mapped to switching between the various
``information-bearing'' states of a computer as it executes an algorithm.
This abstraction of the concept of computation from its material basis, where we can fruitfully
apply it to understand biological transformations, owes its origin to
the British mathematical genius Alan Turing~\cite{Saunders1993}.

Indeed, once the structure of the genetic material, viz., DNA, was shown to be
essentially a one-dimensional string of four constituent molecules 
strung together one after another, the paradigm that came to the fore
involved an abstract theoretical representation of computing
termed the Turing machine. Turing had proposed this in 1936 as part of
a solution to one of the most outstanding open mathematical problems
of the twentieth century: is it possible to come up with a systematic
procedure by which one can decide if any mathematical statement can be proved from a given set of axioms, using a finite number of logical manipulations?
Or in other words, can mathematics be mechanized? Turing proved that this
is not possible by employing an abstract machine (inevitably named as the
``Turing machine'' by others) comprising a head that could read, write
and move along an infinite tape divided into squares on which symbols are
(or can be) inscribed. A series of rules (the program) decides what the
head will do according to its current state (which can change on reading
a particular symbol on the tape) and the symbol on the tape it is scanning
at a given moment. Turing showed that the program itself can be supplied by
the tape, so that in essence the machine can simulate any other possible 
machine. As a result of this, he is considered one of the founding fathers
of the modern digital computer along with John von Neumann~\cite{AlHashimi2023}.
Later developments have shown that the Turing Machine is an apt analogy for 
the way information in DNA or RNA is processed -- it can be seen as either 
instruction to be acted upon (Ribosome as the TM) or data to be copied (DNA
Polymerase or RNA Polymerase as the TM).

It is intriguing that Turing's ``doppelg\"{a}nger'', von Neumann, had also invoked an abstract model for computation when he turned his attention to
problems in biology, specifically, self-replication and brain functioning.
Aspray~\cite{Aspray1990} notes that while von Neumann did not have an 
explicit definition for automata, from the way he used it in his lectures
and writing, one can infer that he had in mind a system that processes
information to implement implement self-regulation (in other words,
control its own behavior). In an early attempt at quantitatively studying
emergence, von Neumann suggested that a complex system can be investigated
at two scales: (i) at the level of the individual components of the system, be they neurons in the case of brains or integrated circuits in the case of computers, and (ii) at the level of the system where these components 
are interacting with each other -- which he considered as the domain
of automata theory\footnote{This 2-step approach to understanding how
a complex system works reminds one of Sydney Brenner's program to understand
how the genes of an organism specify its behavior by decomposing the
question into two parts, the first concerned with
how genes determine the ``wiring diagram'' of the nervous system,
and the second with how the collective dynamics of the nervous system
leads to specific behavioral output\cite{Brenner1973}.}.
In particular, to show that machines are, in principle, capable
of self-replication, von Neumann proposed a computational paradigm called
cellular automata, involving a grid of cells -- where each cell could be in
a finite number of states. The transition between these states are
governed by a set of rules that decide the state of a cell in the immediate
future (the output) based on information about
its present state and that of its neighboring cells (the input). 
Computation in such systems can thus be envisaged as the discrete time
evolution of a spatially extended dynamical system, allowing a formal mapping of biological phenomena to information processing.

\begin{figure}[htbp]
   \includegraphics*[width=\columnwidth]{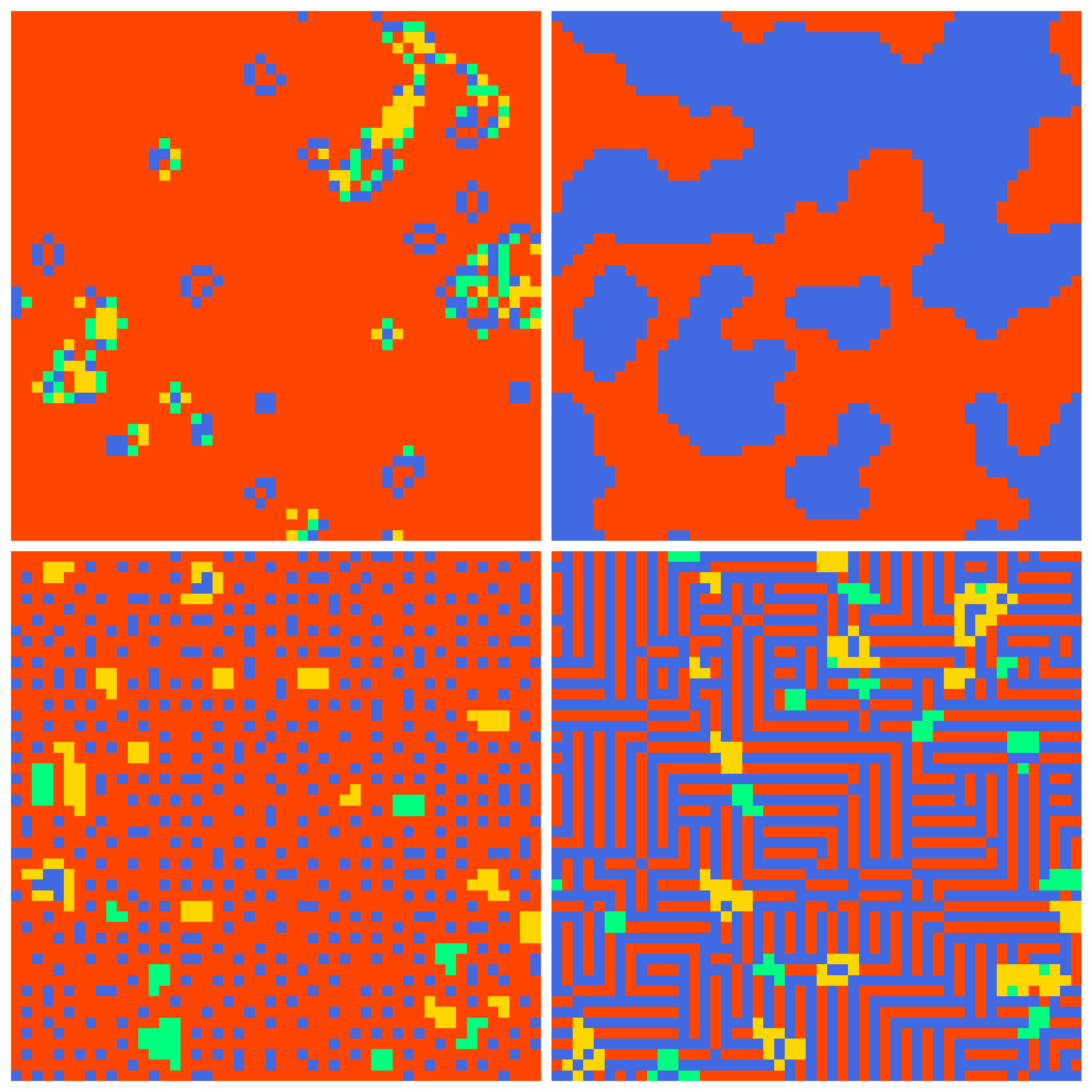}
   \caption{Characteristic dynamic patterns obtained using the game of life (top left), as well as other $2$-D cellular automata that obey similar rules but with different choices of the threshold values. The colors represent the change in the state of each sites between the current and previous iterations, viz. blue ($1\rightarrow 1$), red ($0\rightarrow 0$), yellow ($1\rightarrow 0$), green ($0\rightarrow 1$). We observe that such rules can yield a wide range of complex spatiotemporal patterns.}
      \vspace*{-2mm}
   \label{fig:Lifesnaps}
\end{figure}
Possibly the most well-known cellular automata is the so-called
``Game of Life'' invented by the mathematician John Horton Conway in 1970,
popularized by Martin Gardner through his ``Mathematical Games''
columns in \textit{Scientific American} magazine~\cite{Gardner1983}.
The cells in this automata are all binary-state devices (they can be referred
as ``0" or ``dead'' and ``1" or alive), occupying a two-dimensional square lattice. The state of the cells in the Moore neighborhood of the central
cell, defined as the eight cells that surround it, provide the input
to the rules, which specify that if a ``live'' cell (i.e., in state 1)
is surrounded by 2 or 3 ``live'' cells, it will ``live'' in the next
iteration, else it ``dies'' (i.e., switches to state 0). On the other hand,
if a ``dead'' cell is surrounded by exactly 3 ``live'' cells, it comes
``alive'' (i.e., switches to state 1) in the next iteration, else it remains ``dead''. The repeated application of these deceptively simple rules lead 
to unexpected complexity if a sufficiently large lattice is considered (Fig.~\ref{fig:Lifesnaps}),
with the appearance of localized defects called ``gliders'' traveling through space retaining their configuration and even more complicated structures called ``glider guns'' that generate 
``gliders'' after every 30 iterations. Indeed, it has been shown 
that not only can the automata be used to construct logic gates for performing
specific types of computation, but it
is in fact capable of universal computation in the sense of Turing -- i.e., it 
can be ``programmed'' to simulate any other automata.
One can ask whether such capabilities are extremely rare among cellular
automata, and thus will be unlikely to emerge through self-organization
subject to fluctuations. Fig.~\ref{fig:Lifesnaps} compares ``Life'' with
other two-dimensional automata that follow somewhat different rules
which, although less general than `Life'', nevertheless exhibit non-trivial
spatio-temporal patterns~\cite{Dutta2025}.
Readers familiar with statistical mechanics will note that the 
famous Lenz-Ising model having ferromagnetic interactions can be seen as a  two-dimensional
cellular automata in which each cell follows a majority rule with respect
to its neighborhood (i.e., the
future state of a cell is the same as that of the majority of its neighbors).

One of the reasons why approaches based on automata theory or more generally, utilizing the concept of computation, have not been widely adapted for
explaining biological processes is perhaps because the resulting models
typically evolve through a discrete set of states over discretized space-time.
While this makes it easy to connect with mainstream models of computing,
conventionally we describe natural processes using continuous variables
and over a space-time continuum. Thus, to be useful for analyzing living
systems, the concepts we have talked about above, such as symmetry breaking,
should be also seen in continuum descriptions of biological phenomena.
Intriguingly, Turing, one of the pioneers in creating the vocabulary
for computing science, is also responsible for one of the most
influential continuum models for describing self-organized pattern formation
that was motivated by his interest in symmetry-breaking during 
morphogenesis, a phenomenon that we discussed earlier. In a pioneering paper written shortly before his death~\cite{Turing1952}, Turing outlined the
basic requirements of a system to exhibit a transition from a homogeneous
equilibrium to one characterized by spatially periodic variations in the
state variable (e.g., molecular characterizations) typically seen as
spots or stripes. In the simplest setting all this requires is the
existence of two species of molecules (referred to as ``morphogens'' by Turing), one (the `activator') promoting the production of itself as well as the other molecule (the `inhibitor'), which as its name implies, tends
to suppress the production of the activator. If one neglects space, this
will result in the system attaining an equilibrium. However, in a spatially
extended system where neighboring regions are coupled by molecular transport
as a result of diffusion, one may observe that diffusion, acting in conjunction with the reaction between the two species of molecules, instead of smoothing out concentration heterogeneities actually destabilizes the
homogeneous equilibrium. Such a `Turing instability' occurs when the
inhibitor diffuses much faster than the activator, leading to 
amplification of any initial minute fluctuations in the concentrations
of the molecules. Eventually, the process gives rise to a time-invariant
spatially periodic pattern whose wavelength is related to the diffusion
rates of the two types of molecules.

\begin{figure}[htbp]
   \includegraphics*[width=\columnwidth]{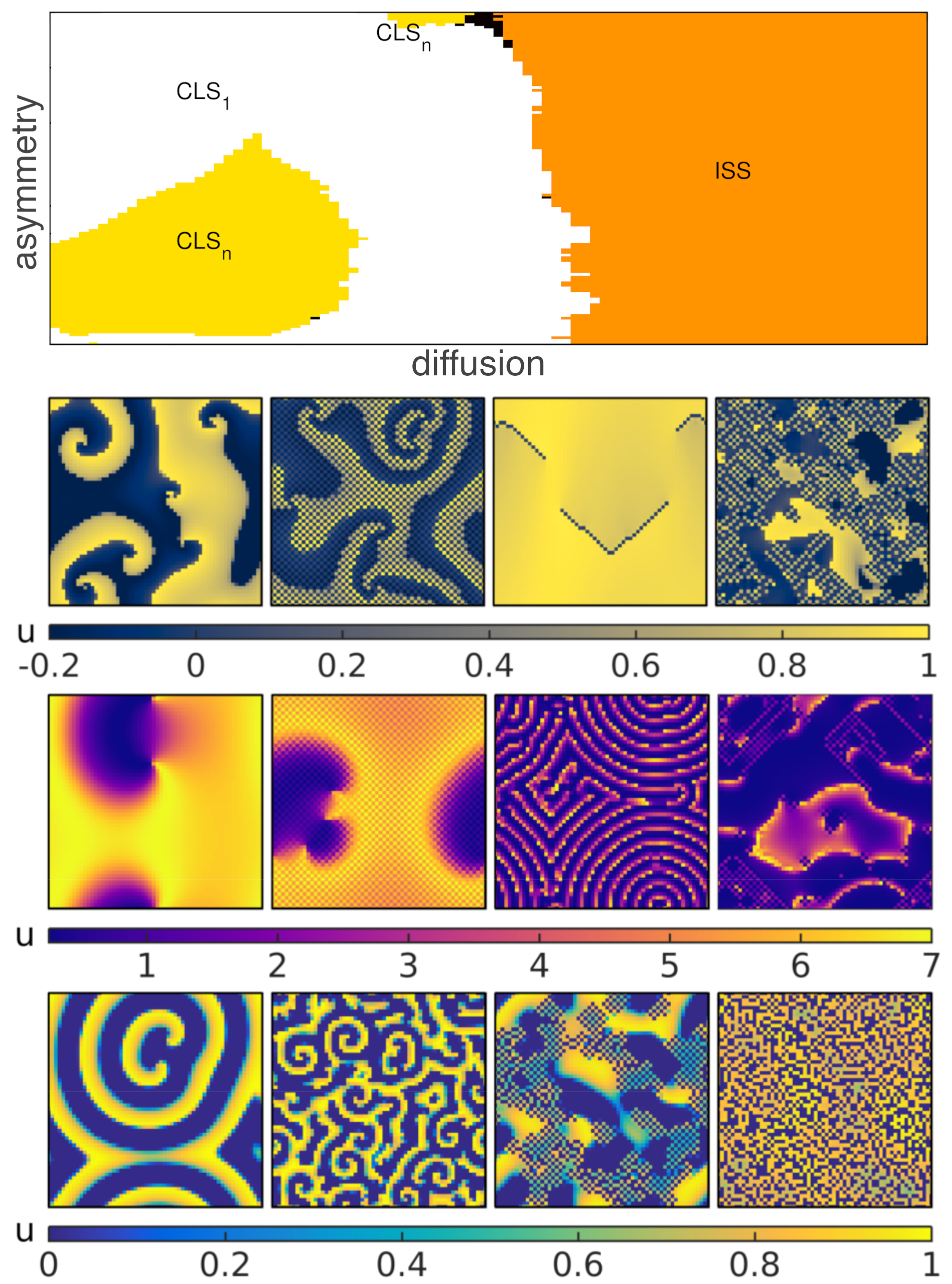}
   \caption{[top row] Regimes of collective dynamical patterns in a system of globally coupled relaxation oscillators over a range of diffusion strengths and upon varying the asymmetry of the limit cycle for individual oscillators (panels adapted from Ref.~\cite{Janaki2019}). We broadly observe two fundamental dynamical regimes, viz. Cluster Synchronization (CLS) and Inhomogeneous Steady State (ISS) at low and high diffusion strengths, respectively. The subscript $n$ in CLS$_{n}$ denote the number of clusters that the oscillators are grouped into. [rows 2-4] Snapshots of collective patterns observed for systems of different types of relaxation oscillators (row 2: the FitzHugh-Nagumo model, row 3: the Brusselator model, and row 4: the Rosenzweig-MacArthur model) diffusively coupled on a lattice, displaying a sample of the variety of complex dynamical behaviors that such systems are capable of exhibiting.}
      \vspace*{-2mm}
   \label{fig:LatInhib}
\end{figure}
While Turing patterns are perhaps the most widely studied framework for diffusion-mediated pattern formation, there exist other paradigms that are capable of explaining the emergence of non-trivial collective dynamics in a variety of chemical, biological and ecological contexts. For instance, a wide array of spatiotemporal patterns can be observed in systems of diffusively coupled relaxation oscillators, which comprise activatory and inhibitory components that operate at fast and slow time-scales, respectively. 
Before describing such systems, let us digress briefly into the 
investigation of time-varying or dynamic patterns in biology. While
temporal oscillations are ubiquitous in biology -- one needs to only think
of the regular cardiac muscle contractions that allow the heart to
pump blood throughout the body or the daily sleep-wake cycle coordinated
by the circadian rhythm -- strangely, the quantitative investigation
of biological oscillators is of relatively recent vintage.

The shift of focus to describing oscillatory behavior in organisms
can be linked to the realization of the physiologically crucial regulatory role played by feedback inhibition, exemplified by enzymatic activity being repressed by one of the metabolic products of the reaction that the
enzyme promotes. While several examples of feedback inhibition had
been known by the early 1960s~\cite{Jacob1961},
the dynamical implication of this mechanism were not yet formally investigated till Brian Goodwin, among others, showed using mathematical modeling how such systems can be driven to oscillatory state upon varying parameters so as to cross a critical value~\cite{Gonze2021}.
Crucially, these oscillations were not sinusoidal in nature (such as
those exhibited by harmonic oscillators, e.g., the pendulum, that are familiar to physicists) but exhibited a variety of forms, including
large-amplitude spikes having a brief duration separated by long intervals
of quiescence -- a pattern observed when recording a series of action potentials from a neuron driven by a steady external current~\cite{Sinha2015}.
These \textit{relaxation oscillations}, so-called because their activity 
is an outcome of slow buildup followed by a rapid release (relaxation), 
are not only observed in the living world across scales from cell to ecosystems, but had also been studied in circuits as early as the 1920s
by the Dutch electrical engineer and physicist Balthasar van der Pol.
However, their widespread appearance in physiological processes, such as biochemical reactions~\cite{Goldbeter1996}, only began to be appreciated from the 1960s.

The presence of a multiplicity of oscillators -- with their
frequencies distributed over a physiologically allowed range -- in living systems, naturally bring about the question of
the phase relationship between these oscillators. That interacting oscillators
can, starting from arbitrary phases, coordinate their phase difference,
has been known from at least the 17th century. In 1665, Christian Huyghens
had reported this phenomenon, when observing two pendulum clocks that were mechanically coupled via their vibrations transmitted along a beam, synchronized their motion so as to have the same period but with the two pendula swinging in opposite directions (anti-phase synchronization).
In the living world such synchronization of relaxation oscillators can be
seen in the context of certain types of fireflies timing their flashing
to occur simultaneously, pacemaker cells in the heart coordinating their
activation, beta-cells in the islets of Langerhans located in the pancreas getting excited together which enables insulin secretion, etc. Such
collective behavior often has critical physiological implications -- e.g.,
the undesirable synchronization of neural firing activity seen in epilepsy
or the loss of coordination in activation of cardiac myocytes during
heart arrhythmias -- and therefore has been the subject of much theoretical analysis~\cite{Winfree1967}. However, for analytical simplicity, such studies
have often described the individual oscillators using only a phase description, that ignores the fundamentally non-sinusoidal nature of
relaxation oscillations. Furthermore, the coupling is often assumed to
be linear at small values of phase-differences. Such assumptions lead
typically to results that imply increasing coupling strength will always
result in monotonic diminishing of the phase difference -- with complete
phase synchronization being achieved above some critical value of 
coupling. However, the fundamentally nonlinear nature of biological
interactions, e.g., that between oscillators describing local activity in
a brain region comprising many excitatory and inhibitory neurons, may
result in dramatically different results -- such as, increasing coupling
leading to desynchronization~\cite{Tripathi2020}.

%
%
%
%
%

%
\begin{figure}[tbp]
   \includegraphics*[width=\columnwidth]{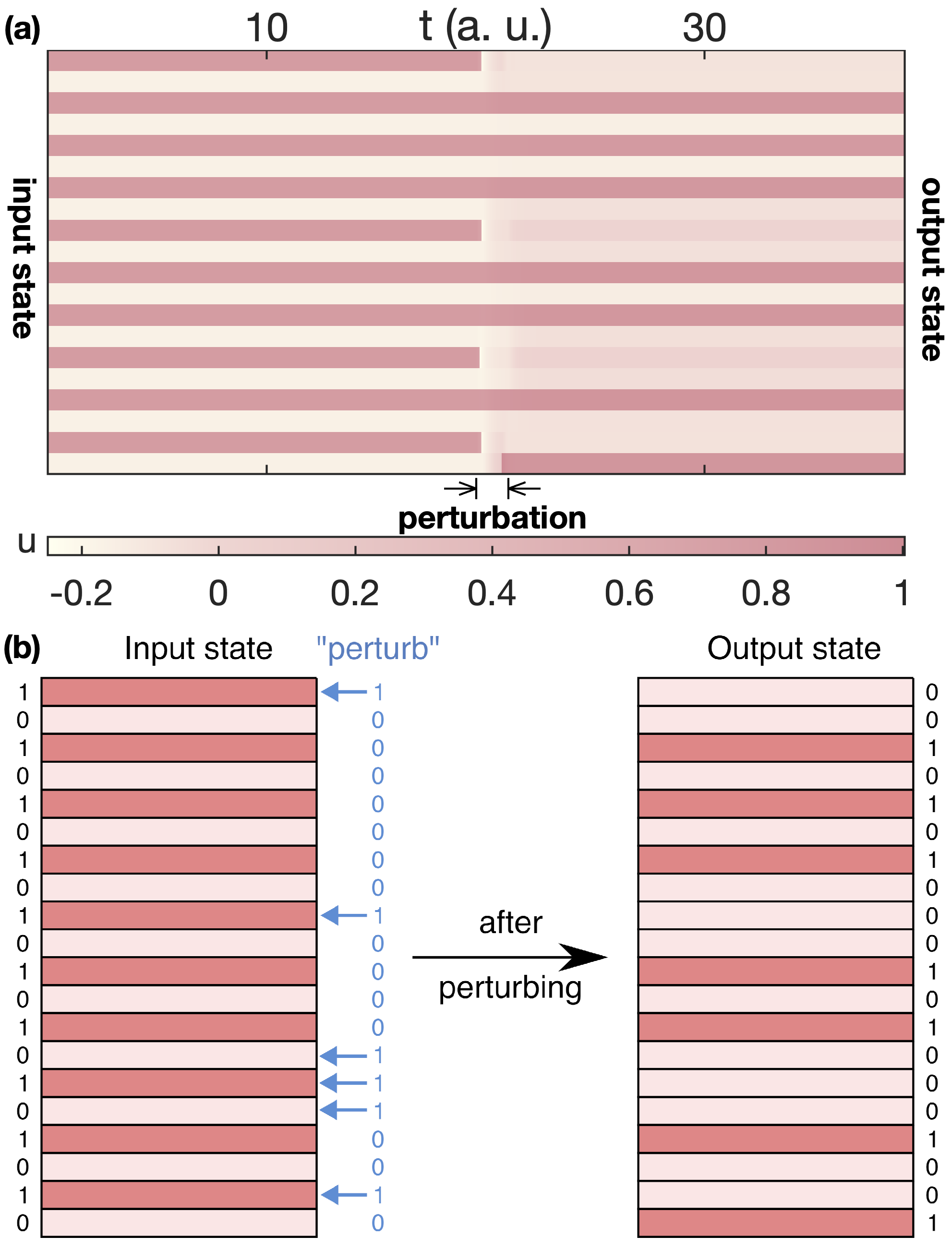}
   \caption{(a) Effect of providing a transient perturbation to a system of relaxation oscillators coupled to their neighbors via diffusion, in a parameter regime where they exhibit a collective dynamical state characterized by the arrest of all oscillators, viz. SPOD (spatially patterned oscillator death). A small brief perturbation causes the system to reorganize and converge to a different SPOD state. (b) As the individual oscillators in a SPOD state can assume either high or low values, separated by a significant amount, it is possible to represent such states as binary strings. Similarly, the perturbation can also be thought of as a binary string of logical variables that represent whether an individual site is perturbed (1) or not (0). Hence, this perturbation-induced transition can be viewed as a transformation involving binary strings, allowing us to probe the inherent computation being performed by the system.}
      \vspace*{-2mm}
   \label{fig:SPOD}
\end{figure}
%
Interactions also lead to a much richer variety of collective
behavior than just simple phase synchronization.
Microfluidic experiments on spatially extended arrays comprising beads of chemical relaxation oscillators have demonstrated that different types of synchronization behaviors can arise via the diffusion of the inhibitory component~\cite{Toiya2008}. It was later shown that this phenomenon can be modeled in a system of generic relaxation oscillators, viz. the FitzHugh-Nagumo (FHN) model, diffusively coupled through the inactivation variable, making it possible to probe the roles of diffusion strength and the asymmetry of the limit cycle on the nature of the emergent collective dynamics~\cite{Singh2013}. Given the simplicity of the two-variable FHN model, a natural question is to what extent the predicted results might vary across other types of relaxation oscillators.

\begin{figure*}[htbp]
   \includegraphics[width=\textwidth]{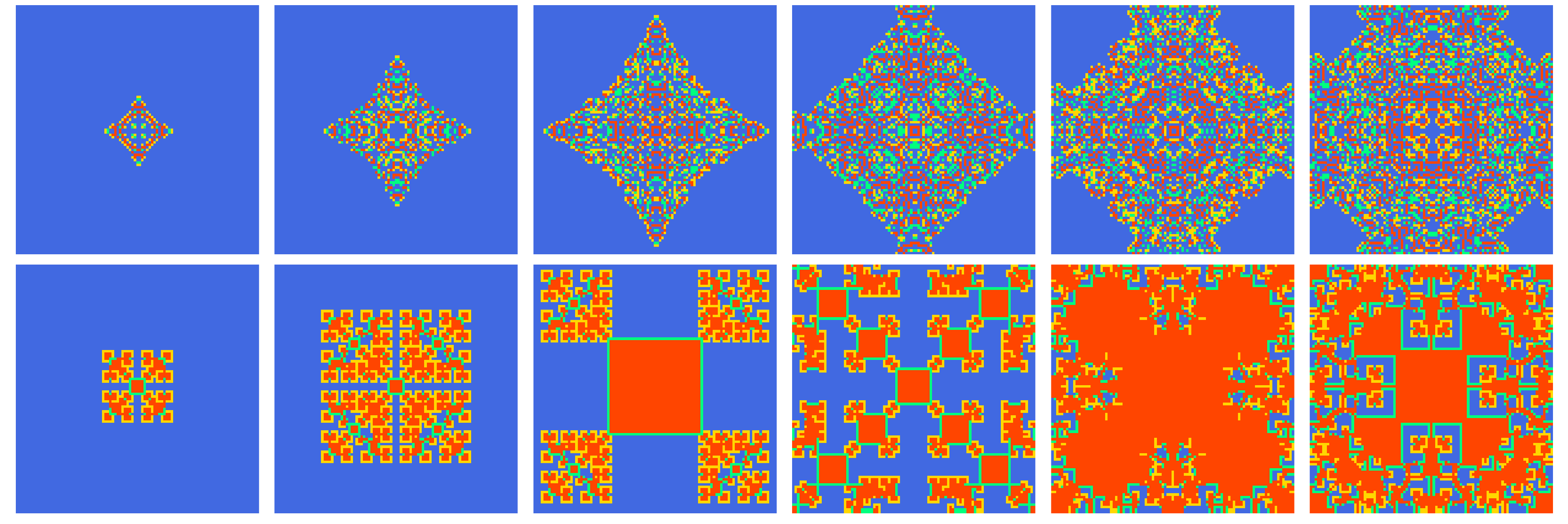}
   \caption{Comparison of the spatiotemporal evolution of a pattern obtained using a $2$-D cellular automata with rules similar to the game of life, using a different choice of threshold values, with the carpet-like  pattern obtained with the spatial Prisoner's Dilemma game \cite{Nowak1992}. The latter comprises a system of agents on a lattice playing a two-player game with each of their neighbors, accumulating the payoffs from all games and then adapting the action of the most successful neighbor in the next iteration. The colors represent the change in the state of each site between two successive iterations, viz. blue ($1\rightarrow 1$ or $C\rightarrow C$), red ($0\rightarrow 0$ or $D\rightarrow D$), yellow ($1\rightarrow 0$ or $C\rightarrow D$), green ($0\rightarrow 1$ or $D\rightarrow C$).}
      \vspace*{-2mm}
   \label{fig:LifevsPD}
\end{figure*}
\begin{figure}[htbp!]
   \includegraphics*[width=\columnwidth]{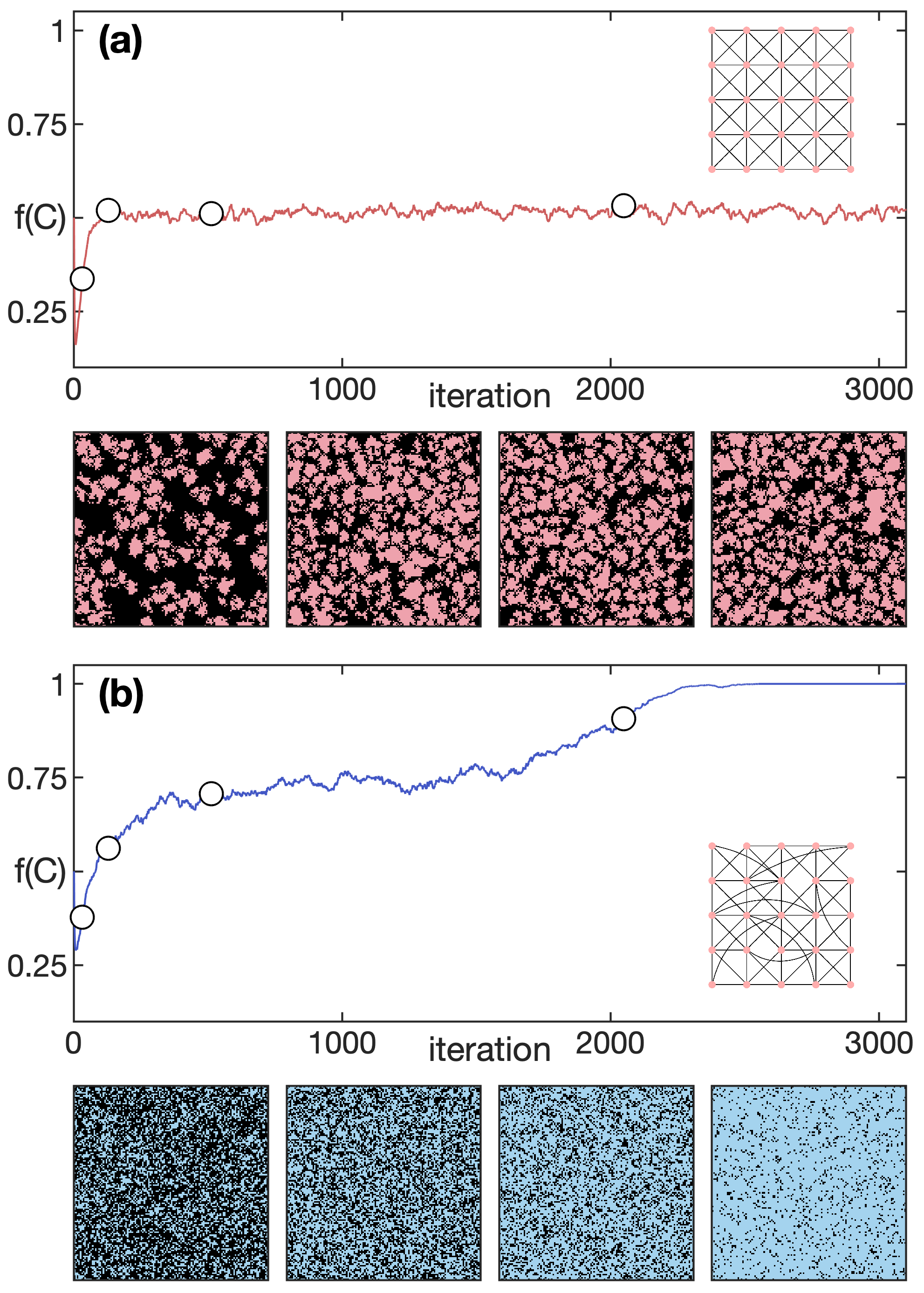}
   \caption{Comparison of collective dynamics of a system of agents playing an iterated synchronous two-player game with each of their neighbors on a network, for different probabilities $p$ that each link is rewired using the Watts-Strogatz protocol~\cite{Watts1998}. (a) Dynamics of a system of $N=16384$ agents arranged on a regular square lattice ($p=0$), where each site is connected to $8$ nearest neighbors (see inset). The system is initialized such that $50\%$ of the agents are randomly selected to cooperate at the first time step. Every agent plays a game with each of their neighbors, accumulates a payoff from each of the games, and synchronously update their actions using fermi rule with a finite temperature. The fraction of cooperators $f(C)$ is displayed as a function of iteration number, and snapshots of the lattice are displayed below at different time points (indicated in the top panel as white circles). The black sites in the four snapshots represent defectors, while the colored sites represent cooperators. (b) Dynamics of a system of $N=16384$ agents on a network obtained by randomly rewiring each link of the regular square lattice with probability $p=0.8$ (see inset). The snapshots at the iterations indicated by white circles are shown below, where once again black sites represent defectors while colored sites represent cooperators. Note here that clusters are not apparent because there exist numerous long-range connections that are not displayed here.}
      \vspace*{-2mm}
   \label{fig:PD}
\end{figure}
%
It was shown in Ref.~\cite{Janaki2019} that qualitatively similar spatiotemporal behavior is observed in spatially extended arrays of relaxation oscillators where each oscillator is coupled to its neighbors via diffusion of the inhibitory component of its dynamics, regardless of the specific details of the reaction process. In particular, systems as diverse as the Brusselator model for autocatalytic chemical reactions, the Rosenzweig-MacArthur model of predator-prey dynamics and a model of the cell-cycle all display the same types of dynamics observed with the FHN model. These include complex synchronization patterns, ``chimera'' patterns comprising oscillating and non-oscillating elements, and patterns in which the activities of all oscillators are arrested, which we refer to as Spatially Patterned Oscillator Death (SPOD). Despite the complexity of observed patterns, it was shown that all of the observed dynamics can be categorized into two classes, viz. Cluster Synchronization (CLS) and Inhomogeneous Steady State (ISS), which are typically seen at low and high diffusion strengths, respectively (see Fig.~\ref{fig:LatInhib}). The fact that the broad nature of the collective dynamics is more or less independent of the kinetics of the associated reactions suggests that such states are ubiquitous in natural systems. This immediately raises the following question: do transitions between these states encode vital information that may underlie essential functions of the system?

If computation is defined as a rule-based transformation of an input state to an output, one can then view dynamical systems as performing a sequence of logical operations. Indeed, this connection is well-established in cellular automata, several of which have been shown to be capable of universal computation. However, it has been challenging to attain an unambiguous logic gate representation in systems with continuous variables~\cite{Chatzinikolaou2022}. It has been observed that systems of relaxation oscillators, coupled via diffusion of only the inhibitory component, can exhibit numerous complex collective patterns that, upon perturbation, can be transformed to a different pattern. This property was used as the basis for a proposition for implementing computation using a class of dynamical patterns observed in such systems of oscillators~\cite{MenonIEEE}, where it was shown that it is in principle possible to view the system as a logic gate. However, although such transformations are theoretically feasible, such a framework would in practice be challenging to implement in experiments.

To this end, an alternative framework has been proposed by us, focusing on a temporally invariant pattern observed in the same system, viz. the SPOD state. Such a state is characterized by oscillators arrested at either high or low values of the associated variables, and hence they can unambiguously be represented as a binary string. When a SPOD state receives a brief perturbation, where the choice of whether (1) or not (0) to perturb a site also can be thought of as a binary string, the system undergoes a reconfiguration before resting to another SPOD state (or returning back to the same SPOD state). Thus, this entire process can be interpreted as a computation (see Fig.~\ref{fig:SPOD}), and by studying the nature of transformations exhibited by the system, it is possible to determine the types of logical operations performed by the system. It was found~\cite{Shamim} that the rules governing the possible transitions between SPOD states emerge from an underlying energy landscape, which is shaped by both the local and global stabilities of the individual states. In particular, it was found that the size of the basin of attraction of each state displayed a clear dependence on its energy, defined using a mapping to an anti-ferromagnetic spin chain. Although spin models are unreasonably effective (to paraphrase Wigner's famous statement) for studying programmed decision-making in natural systems, and oscillator models provide a suitable framework for describing the dynamics emerging from the inherent rhythmicity of several biological processes, such approaches cannot capture the inherent autonomy of many organisms. Such behavior can only be described via a full account of the costs and benefits associated with each outcome, through the framework of game theory, allowing one to explore strategies employed by the organism in the attempts to to achieve an ``optimal'' decision.

A fundamental distinction between living organisms and the typical 
components used in physics-inspired models of them is that the former
have ``agency'' (hence the term ``agents'' used in game-theoretic
literature). While non-living entities would be described in a
framework where their past states (and that of their environment) directly
dictate their future, agents are goal-directed and hence, their present
behavior is typically conditioned by their perception of the future.
While such perceptions are constructed from past experiences, it is 
also shaped by intrinsic properties (such as attitudes or beliefs) that
may not be reducible to a simple function of past events. 
Interaction between such agents can often lead to counter-intuitive
collective actions as a result of this inherent complexity.
Thus, in the event of an epidemic where the disease can be controlled by
an available vaccine, we may see many rational agents (i.e., those 
who do not wish to suffer by being infected with the disease) 
nevertheless deciding against 
getting vaccinated. This unexpected scenario arises when agents decide
independently that as long as enough number of their neighbors
have got themselves vaccinated, they will be protected by virtue of being
surrounded by a screen of individuals who cannot be affected by the
disease-causing pathogens. Thus, by trying to free-ride on the
efforts of others who have got themselves vaccinated, they wish
to reap the benefit of vaccination without bearing any cost (which
need not be monetary) of getting vaccinated themselves. As this outcome
depends on the perception of each agent of their relative risk of being
infected versus the temptation to free-ride on others' efforts, the
collective action may be controlled by tuning their access to information
about the local (as opposed to global) incidence of the disease~\cite{Sharma2019} (Fig.~\ref{fig:fig1}).

An enduring question that has long interested evolutionary biologists, social psychologists, economists and physicists alike is why cooperation is ubiquitous in nature. One of the many possible mechanisms proposed to explain the emergence of cooperation is ``network reciprocity'', wherein agents interacting via social networks form cooperative clusters to achieve a sufficient net payoff, and hence prevent the entry of defectors. This is a generalization of the concept of ``spatial reciprocity'', shown by Martin Nowak and Robert M. May in a system of agents that interact with their neighbors on a square lattice~\cite{Nowak1992}. They demonstrated that even a simple update rule, viz. imitating the action of the best performing neighbor, was sufficient to ensure that cooperation persists in the system. 
Subsequently, there have been numerous studies investigating the dependence of cooperation on the choice of update rule and various aspects of network topology. We would like to mention at this point that the spatial patterns
obtained in spatial games such as those proposed by Nowak and May are echoed
by two-dimensional cellular automata that we have recently explored that
use rules somewhat different from Conway's ``Game of Life'' (Fig.~\ref{fig:LifevsPD}).

In Ref.~\cite{Menon2018} it was examined how the emergent collective dynamics in a group of agents is jointly affected by stochasticity at the level of individual decision-making and the topology of the network over which agents interact. It was observed that upon interpolating between a regular lattice and a random network with probability $p$, using the ``small world'' network construction procedure proposed by Duncan Watts and Steven H Strogatz~\cite{Watts1998}, one finds the emergence of a regime characterized by the absorbing state of pure cooperation (All-C), i.e. every agent in the system cooperates (Fig.~\ref{fig:PD}). In particular, it was seen that for a given value of noise and temptation, upon varying $p$ one observes a topological-driven transition from a fluctuating state (wherein all nodes permanently fluctuate between cooperation and defection) to an All-C state.

While the transitions in the collective behavior described for agents using strategies to compete with others in order to maximize their payoffs in a game may superficially seem similar to those shown by systems of interacting spins or coupled oscillators, we would like to point out a fundamental difference. Unlike the latter, the behavior of agents in a game cannot be seen in terms of optimizing a function that is common to all of them. Thus, whereas spins are all trying to minimize a common free energy function,
and the oscillators are trying to maximize, say, their coherence with each
other, an agent in trying to maximize her payoff would often have to
choose an action that will lead to a worse payoff for other agents.
While one can try to write down an effective energy function \textit{a posteriori} for describing the collective state that the agents converge to
according to the solution framework adopted, it does not appear to be possible
to define such an energy function from first principles, owing to the
opposing nature of the goals of individual agents. This poses an intriguing
challenge to statistical physicists -- viz., how to analyze a system
of interacting entities for which a common function to be optimized does
not exist? One hopes that in trying to devise such a framework, we will be
able to have a glimpse of the new
kind of physics that may be required to quantitatively investigate systems 
of interacting goal-directed components.\\

\noindent 
\textbf{\textit{Concluding Thoughts}}

Although biology, as it is commonly practiced, is typically not centered around
`theory', biologists nevertheless rely on informal mental `models' to make sense
of their observations. The trouble is that every scientist may have their own
personal models that may not be fully compatible with others, and hence a
coherent unifying framework remains elusive. The situation is comparable to the
lack of a common accepted standard in cell phone charger design, so that one
often ends up with another distinct charger in their collection every time they
switch to a new phone brand. What theoretical biology needs is a generally
accepted framework in which to organize empirical observations. As Sydney Brenner noted in
his Nobel Prize speech, the problem with biology is that it does not have a
theoretical edifice that can help make sense of its facts~\cite{Brenner2002}.
To see a parallel, one can invoke the period of late 16th century when the Danish
astronomer Tycho Brahe had a vast number of celestial observations, but each of
them were isolated data points (as each observation was unique) -- an enormous
smorgasbord of factoids without any overall coherent picture about how
celestial bodies move. It took Johannes Kepler, the German polymath,
to condense all of these into a simple set of principles (viz., the three laws
of planetary motion named after him) from which one can, in principle,
predict each of the original observations. In this sense, the evolution
of theoretical physics can be seen as an endeavor to arrive at a minimal description
(which can, in principle, be quantified in terms of bits) 
that would allow one to recover the entire set of
empirical observations recorded. In the absence of such a description,
each observation would appear to be independent of the others, and would
need to be individually memorized requiring an enormous
amount of memory storage. 

However, theory is more than just simply a means
of lossless data compression -- it should also be able to generalize,
so that it can predict novel phenomena that have not yet been observed.
If these predictions do not match future observations, it implies that the
function used to fit the data is too specific to previous observations (what
one would call ``overfitting'' of data, in today's parlance), and is
consequently unable to generalize properly. The challenge is to come
up with an appropriate fitting function (i.e., a theoretical edifice)
from an extremely limited sample of observations. As the state-space
of a sufficiently complex systems would be inevitably very high-dimensional,
the number of possible observable combinations that can occur would
vastly outnumber the observations that can be made over any reasonable 
duration. This makes the problem severely under-constrained and a large
number of possible functions can be advanced to fit the observations available
to date.

Given the non-uniqueness of possible theories that are consistent with all known facts, it may seem that any choice between them would be rather
arbitrary. However, the aggregative nature of the scientific enterprise
means that a framework with certain key ideas are established early on, 
such as the concepts of energy and entropy in the case of classical theoretical physics,
that later developments or modifications try to fit in with. For
theoretical biology, the key concept may be that of \textit{information} and its associated
operations, such as (a)~how it can be communicated across time 
(from an individual to its offspring over successive generations)
or space (across cells in an organism), (b)~how it can be coded so that it can be exchanged with minimal loss (the deterioration of information
content being an outcome of noise or fluctuations in the
channels over which information is communicate), and (c)~how it can be
transformed into instructions for controlling processes vital for the
survival and reproduction of cells and organisms. This can be expressed
in terms of an acronym that is well-known in an entirely different context,
viz., $C^3 I$, or Control, Communication, Coding and Information.

The attentive reader may well note at this point that the above
basis we suggest for a ``theory'' of biology does not contain the word
`computation', even though that has been a recurrent theme 
throughout our article. That is because we view computation as the totality
of transformations to which information is subjected in all aspects
of living processes.
In fact, in our view, it's not a stretch to suggest theoretical computer science as a template on which to formally model a science of theoretical biology. In this, we are however, only reasserting a perspective that
had been explicated in detail by the computer scientist David Marr more than four
decades back. As Marr noted at the beginning of the book~\cite{Marr1982} that he wrote just before his untimely death, questions about a system as complex
as the brain (or one can add, living systems, in general)
can be understood at multiple levels of abstraction or generality.
In particular, he proposed a framework comprising three levels of
increasing abstraction: (i)~the implementation or the hardware level,
where one asks how the computations that are necessary for achieving a goal
are physically realized, (ii)~the algorithmic level, where one figures
out the series of steps the computation requires, and (iii)~the level
of goals or strategies, which is about finding out what problem(s) the
system is trying to solve and why it needs to solve it. While the field of computational
biology is more concerned with describing biological systems at the
implementation or hardware level, we believe that the goal of a 
theory of biology should be to understand the principles of living
systems at the level of strategies and algorithms. Poised as we are at a time
when natural and artificial forms of life are less distinct than they ever have been, it seems imperative to take a more generalized view of
how life works, or indeed, what are all the possible ways in which 
life (however we choose to define it) may work.\\

\noindent 
\textbf{\textit{Acknowledgments}}

We would like to thank Ananta Dutta and Richa Tripathi for assisting with the preparation of the figures, and Sumithra 
Surendralal for several helpful comments during the preparation of the
manuscript. We also acknowledge the
valuable
discussions we have had over the years with several present and former members of the Complex Systems \& Data Science (CSDS) group at IMSc Chennai. SNM has been supported by the Center of Excellence in Complex Systems and Data Science, funded by the Department of Atomic Energy, Government of India.


\end{document}